\documentclass[
showkeys,12pt,
preprint,preprintnumbers,nofootinbib,
groupedaddress,superscriptaddress,amsmath,amssymb]{revtex4}
\usepackage{graphicx}
\usepackage{dcolumn}
\usepackage{bm}
\usepackage{amssymb}
\usepackage{amsmath}
\usepackage{epsfig}    
\usepackage{color}
\usepackage{slashed}
\usepackage{hhline}

\def\be{\begin{equation}}
\def\ee{\end{equation}}
\newcommand{\bea}{\begin{eqnarray}}
\newcommand{\eea}{\end{eqnarray}}
\newcommand{\nn}{\nonumber}

\numberwithin{equation}{section}

\begin{document}

\title{A Three-loop Neutrino Mass Model with a Colored Triplet Scalar}
\preprint{KIAS-P16080}
\author{Kingman Cheung}
\email{cheung@phys.nthu.edu.tw}
\affiliation{Physics Division, National Center for Theoretical Sciences, 
Hsinchu, Taiwan 300}
\affiliation{Department of Physics, National Tsing Hua University, 
Hsinchu 300, Taiwan}
\affiliation{Division of Quantum Phases and Devices, School of Physics, 
Konkuk University, Seoul 143-701, Republic of Korea}

\author{Takaaki Nomura}
\email{nomura@kias.re.kr}
\affiliation{School of Physics, KIAS, Seoul 130-722, Korea}

\author{Hiroshi Okada}
\email{macokada3hiroshi@cts.nthu.edu.tw}
\affiliation{Physics Division, National Center for Theoretical Sciences, 
Hsinchu, Taiwan 300}

\date{\today}

\begin{abstract}
We study a variation of the Krauss-Nasri-Trodden (KNT) model with 
a colored triplet scalar field and a colored singlet scalar field, 
in which we discuss the anomaly coming from $b\to s\mu\bar\mu$, 
fitting to the muon anomalous magnetic moment and the relic density of 
the Majorana-type dark matter candidate, as well as 
satisfying various constraints such as lepton-flavor
violations and flavor-changing neutral currents. Also, we discuss
the direct constraints from the collider searches and the 
possibilities of detecting the new fields at the LHC.
\end{abstract}
\maketitle

\section{Introduction}
The fact that neutrinos have masses perhaps is the only confirmed evidence 
of physics beyond the standard model (SM). Other observations which
also point to physics beyond the SM, such as dark matter, dark energy, 
matter-antimatter asymmetry, are not as convinced as the neutrino mass.
The type of models that can naturally explains the neutrino mass 
is based on loop diagrams, in which the smallness of neutrino mass is 
achieved by suppression of the loop factors. 
Some classic examples are the one-loop Zee model \cite{zee} and 
Ma model \cite{ma}, two-loop Zee-Babu model~\cite{babu}, 
three-loop Krauss-Nasri-Trodden model (KNT) \cite{knt}, etc.

Recently, there was an $2.6\sigma$ anomaly in lepton-universality
measured in the ratio 
$R_K \equiv B(B\to K\mu\mu)/B(B\to K ee) = 0.745 ^{+0.090}_{-0.074} \pm 0.036$ 
by the LHCb \cite{lhcb-2014}. Moreover, sizable deviations 
were recorded in angular distributions of $B \to K^*\mu\mu$ \cite{lhcb-2013}.
The results can be accounted for by a large negative
contribution to the Wilson coefficient $C_9$ of the semileptonic operator
$O_9$, and also contributions to other Wilson coefficients
\cite{Descotes-Genon:2015uva}.

In this work, we study a variation of the original KNT model with
the original scalar fields replaced by a colored $(\mathbf{\bar 3})$ 
$SU(2)_L$-triplet field and a colored $(\mathbf{3})$   $SU(2)_L$-singlet
field (see Table~\ref{tab:2})~\footnote{{Systematic analysis of this model can be found in the last entry of Table 1 in Ref.~\cite{Chen:2014ska}.}}.
The model can accommodate the
neutrino masses and oscillation, and at the same time the model can alleviate
the anomaly in $b\to s \mu \bar \mu$ with additional
contributions to $C_{9}$.
The model also satisfies all the constraints from the 
lepton-flavor violations (LFV),
flavor-changing neutral currents (FCNC), and the oblique parameters.
Finally, we also discuss the direct constraints coming from the LHC searches
and future possibilities of detecting the colored fields of the model.
The most interesting channel at the LHC will be pair production of
$S^a_{-1/3} S^a_{+1/3}$ followed by the decay into two jets plus missing
energies.

{
Here we summarize the differences and improvements over the original 
KNT model and some other related models.
\begin{enumerate}
\item  
The original KNT model accommodates two colorless singly-charged
bosons with different $Z_2$ charges. It always  predicts one massless
neutrino due to its flavor structure, but the muon $g-2$ is always induced with
a negative value~\cite{Cheung:2016ypw}.  
On the other hand, our model has two colored
leptoquark bosons with different $SU(2)_L$ charges (one is a singlet
and the other is a triplet), and three massive neutrinos can be generated.
Although it potentially gives both positive and negative terms to the 
muon $g-2$, the negative one unfortunately overwhelms the
positive one due to the constraint from oblique parameters.

\item 
An alternative KNT model~\cite{Nomura:2016ezz} introduced two 
colored leptoquark bosons, {which are both $SU(2)_L$ singlets,}
and can then explain the neutrino oscillation data and the DM without 
conflicting various severe constraints arising from leptoquarks.
Even though it can give positive values to the muon $g-2$, sizable values 
cannot be obtained because of strong constraints from {
lepton-flavor violations.}
Also, the model does not include any sources to explain the 
anomaly of $B \to K^*\mu\mu$. 
On the other hand, our current model can explain the anomaly of
$B \to K^*\mu\mu$ by modifying  $C_{9(10)}$, which is the only possible way
in the framework of variant KNT models. It is achieved by replacing 
one of the colored $SU(2)_L$ singlet boson by a  colored $SU(2)_L$ triplet
boson.

\item
As another option, we can explain the anomaly $B \to K^*\mu\mu$
via the second class of  modifications in $C'_{9(10)}$.
This is achieved by introducing two colored leptoquarks {in
$SU(2)_L$-doublet and hypercharges $U(1)_Y = \pm1/6$,} instead of the two 
leptoquraks that we  mentioned in  the third or fourth entry of Table 1 
in Ref.~\cite{Chen:2014ska}.
\end{enumerate}

}

This paper is organized as follows.
In Sec.~II, we describe the modified KNT model, the neutrino mass matrix
and the solution to the anomaly in $b\to s \mu \bar \mu$.
In Sec.~III, we discuss various constraints of the model, including
lepton-flavor violations, FCNC's, oblique parameters, and dark matter. 
In Sec. IV, we present the numerical analysis and allowed parameter space,
followed by the discussion on collider phenomenology.
Sec.~IV is devoted for conclusions and discussion.

\section{The Model}

In this section, we describe the model setup, derive the formulas 
for the active neutrino mass matrix, and calculate the contributions
to $b\to s \mu \bar \mu$.

 \begin{widetext}
\begin{center} 
\begin{table}
\begin{tabular}{|c||c|c|c||c|c||c|}\hline\hline  
&\multicolumn{3}{c||}{Quarks} & \multicolumn{2}{c||}{Leptons}& \multicolumn{1}{c|}{Majorana Fermions} \\\hline
& $Q_{L_{q_i}}^a$~ & ~$u_{R_{q_i}}^a$~ & ~$d_{R_{q_i}}^a$ ~ 
& ~$L_{L_{\ell_i}}$~ & ~$e_{R_{\ell_i}}$ ~ & ~$N_{R_{\ell_i}}$ 
\\\hline 
$SU(3)_C$ & $\bm{3}$  & $\bm{3}$  & $\bm{3}$  & $\bm{1}$& $\bm{1}$& $\bm{1}$   \\\hline 
$SU(2)_L$ & $\bm{2}$  & $\bm{1}$  & $\bm{1}$  & $\bm{2}$& $\bm{1}$& $\bm{1}$   \\\hline 
$U(1)_Y$ & $\frac16$ & $\frac23$  & $-\frac{1}{3}$  & $-\frac12$  & $-1$  & $0$\\\hline
$Z_2$ & $+$ & $+$  & $+$ & $+$ & $+$ & $-$  \\ \hline
\end{tabular}
\caption{Field contents of fermions
and their charge assignments under 
$SU(3)_C\times SU(2)_L\times U(1)_Y\times Z_2$, where the 
superscript (subscript) index $a=(1,2,3)$ represents the color,  {and $q_i,\ell_i (i = 1,2,3)$ distinguish the generation of quarks and leptons. }}
\label{tab:1}
\end{table}
\end{center}
\end{widetext}
\begin{table}
\centering {\fontsize{10}{12}
\begin{tabular}{|c||c|c|c|}\hline\hline
&~ $\Phi$  ~&~ $\Delta^a$  ~&~ ${S}^{a}$ \\\hline
$SU(3)_C$ & $\bm{1}$  & $\bar{\bm{ 3}}$ & $\bm{3}$ \\\hline 
$SU(2)_L$ & $\bm{2}$   & $\bm{3}$  & $\bm{1}$  \\\hline 
$U(1)_Y$ & $\frac12$  & $\frac{1}{3}$ & $-\frac{1}{3}$   \\\hline
$Z_2$ & $+$   & $+$  & $-$ \\\hline
\end{tabular}%
} 
\caption{Field contents of bosons
and their charge assignments under  
$SU(3)_C\times SU(2)_L\times U(1)_Y\times Z_2$, 
where the superscript index $a=(1,2,3)$ represents the color. }
\label{tab:2}
\end{table}
\subsection{ Model setup}
We show all the field contents and their charge assignments in 
Table~\ref{tab:1} for the fermionic sector and  Table~\ref{tab:2} for the 
bosonic sector. 
Under this framework, the relevant part of the renormalizable Lagrangian and the Higgs potential are given by
 {
\begin{align}
\label{eq:Yukawa}
-{\cal L}^{}&=
(y_\ell)_{\ell_i \ell_j}\bar L_{L_{\ell_i}} \Phi e_{R_{\ell_j}} + f_{q_i \ell_j}\bar Q^{c a}_{L_{q_i}}(i\sigma_2) \Delta^{ a} L_{L_{\ell_j}}
+g_{\ell_i q_j}\bar N_{R_{\ell_i}} d^{ca}_{R_{q_j}} S^a + M_{N_{\ell_i}} \bar N_{R_{\ell_i}}^c N_{R_{\ell_i}}+{\rm h.c.} ,
\end{align}
\begin{align}
\label{eq:potential}
{\cal V}&=m_\Phi^2 \Phi^\dag \Phi + m_{S}^2 S^{*} S^{}
+ m_{\Delta}^2 {\rm Tr}[\Delta^{\dag a} \Delta^a]  +\lambda_0\left( {\rm Tr}[\Delta^{ a}\Delta^a]S^{b}S^b +{\rm c.c.}\right)
 \nn\\
&  +\lambda'_0\left( {\rm Tr}[\Delta^{ a}\Delta^b]S^{a}S^b +{\rm c.c.}\right) 
 +\lambda''_0\left( {\rm Tr}[\Delta^{ a}\Delta^b]S^{a}S^b +{\rm c.c.}\right)  \nn\\
&+\lambda_{\Phi} |\Phi^\dag \Phi|^2
+\lambda_{S} |S^{*a} S^a|^2 +\lambda_\Delta [{\rm Tr}(\Delta^{\dag a}\Delta^a)]^2 +\lambda_\Delta' {\rm Tr}[(\Delta^{\dag a}\Delta^a)]^2
+\lambda_{\Phi S} (\Phi^\dag \Phi) (S^{*a} S^a) \nn\\
&+ \lambda_{\Phi \Delta} (\Phi^\dag \Phi) {\rm Tr}[\Delta^{\dag a}\Delta^a] 
+ \lambda'_{\Phi \Delta} \sum_{i=1}^3(\Phi^\dag \sigma_i \Phi) {\rm Tr}[\Delta^{\dag a} \sigma_i \Delta^a] 
+ \lambda_{S \Delta}(S^{*a} S^a) {\rm Tr}[\Delta^{\dag b}\Delta^b]
 \nn\\
& + \lambda'_{S \Delta}(S^{*a} S^b) {\rm Tr}[\Delta^{\dag a}\Delta^b]
  + \lambda''_{S \Delta}(S^{*a} S^b) {\rm Tr}[\Delta^{\dag b}\Delta^a],
\end{align}
}
where $\sigma_i$ are the Pauli matrices, {the 
superscript (subscript) index $a=(1,2,3)$ represents the color, and $q_i (i= 1,2,3)$ and $\ell_i (i=1,2,3)$ distinguish the generation of quarks and leptons respectively.}
Each of $\lambda_0$, $\lambda_0'$, and $\lambda_0''$
comes from the contraction of 
$(\bar3\times\bar3)(3\times3)\to(\bar6)\times(6)\to 1$, 
$(\bar3\times3)(\bar3\times3)\to( 1)\times(1)\to 1$ and 
$(\bar3\times3)(\bar3\times3)\to( 8)\times(8)\to 1$. 
Therefore, a color factor of $(6+1+8=)15$ is multiplied to 
$\lambda_0$ as shown in the neutrino mass matrix, 
when we assume that $\lambda_0=\lambda_0'=\lambda_0''$. 

The scalar fields can be parameterized as 
\begin{align}
&\Phi =\left[
\begin{array}{c}
w^+\\
\frac{v+\phi+iz}{\sqrt2}
\end{array}\right],\quad 
\Delta =\left[
\begin{array}{cc}
\frac{\delta_{1/3}}{\sqrt2} & \delta_{4/3} \\
\delta_{-2/3} & -\frac{\delta_{1/3}}{\sqrt2}
\end{array}\right],\quad S^a\equiv S_{-1/3}^a \;,
\label{component}
\end{align}
where {the subscript next to the field represents the electric charge
of the field,} $v= 246$ GeV, 
and $w^\pm$ and $z$ are respectively Nambu-Goldstone boson (NGB) 
which are absorbed by the longitudinal component of $W$ and $Z$ bosons.
Notice that each of the components of $\Delta$ is in the mass eigenstate, 
since there are no mixing terms.

\subsection{Active neutrino mass matrix}
\begin{figure}[tb]
\begin{center}
\includegraphics[width=80mm]{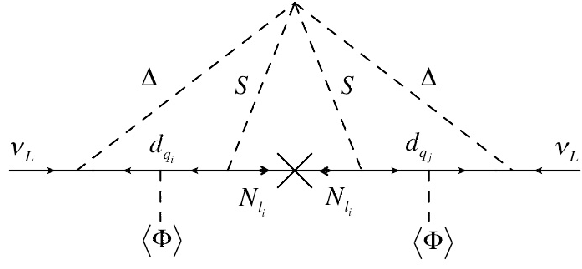}
\caption{{
The Feynman diagram for neutrino mass generation.
}}
\label{fig:diagram}
\end{center}
\end{figure}
The neutrino mass matrix is induced at three-loop level {as shown in Fig.~\ref{fig:diagram}}, and its formula 
is given by 
{
\begin{align}
&{\cal M}_{\nu_{ab}}\approx -\frac{60\lambda_0}{(4\pi)^6 M_{\rm Max}^2}  \sum_{i,j,k=1}^3
f^T_{\ell_a q_i} m_{d_i} g^T_{q_i \ell_k} M_{N_{\ell_k}} g_{\ell_k d_j}m_{d_j} f_{d_j \ell_b}
 F_{3}(r_{N_{\ell_k}}, r_{S_{1/3}}, r_{\delta_{1/3}}) ,\label{Eq:act_mass}\\
&F_{3}(r_{N_{\ell_k}}, r_{S_{1/3}}, r_{\delta_{1/3}})=\int [dx]  \int [dx' ]  \int [dx'']  
\nn\\
&\times
\frac{\delta(1-x-y-z)\delta(1-x'-y'-z')\delta(1-x''-y''-z'')}
{x''(z'^2-z') (y r_{S_{1/3}}+z r_{\delta_{1/3}}) +y'' (z^2-z) (y' r_{S_{1/3}}+z' r_{\delta_{1/3}}) -z'' (z^2-z)(z'^2-z') r_{N_{\ell_k}} },
\end{align}
}
where $m_{d_{ \{1,2,3 \} }} = \{ m_d, m_s, m_b \}$, 
$M_{\rm Max}\equiv$ Max[$M_{N_{\ell_k}},m_{S_{1/3}}, m_{\delta_{1/3}}$], 
$r_f\equiv m_f^2/M_{\rm Max}^2$, 
$[dx]\equiv dxdydz$, and we assume that $m_{\ell} \ll M_{N_{\ell_k}},m_{S_{1/3}}, m_{\delta_{1/3}}$, . 
Note that $F_3(r)$ is given in Ref.~\cite{Cheung:2016ypw},~\footnote{{The typical scale of $F_3$ in our parameter range is 10.}}
and 
$m_{S_{1/3}}$ and $m_{\delta_{1/3}}$ represent the masses 
of $S_{\pm1/3}$ and $\delta_{\pm1/3}$, respectively.

To achieve the numerical analysis of the neutrino oscillation data, we apply a method of Casas-Ibarra parametrization~\cite{Casas:2001sr} to our neutrino mass matrix and its form is explicitly given by~\cite{Nomura:2016ezz}
{
\begin{align}
f &=m_d^{-1}  g^{-1} A^{-1/2} {\cal O} \sqrt{{\cal M}^{diag}_\nu} V_{MNS}^\dag ,\quad
{\rm or}\quad
g =  A^{-1/2} {\cal O} \sqrt{{\cal M}^{diag}_\nu} V_{MNS}^\dag f^{-1} m_d^{-1},\label{eq:cis}
\end{align}
where we define the diagonalization of ${\cal M}_{\nu}$  to be
${\cal M}^{diag}_\nu=V_{MNS}^T {\cal M}_{\nu} V_{MNS}$,
and
\begin{align}
A&\equiv \frac{60\lambda_0  M_{N_{\ell_k}}}{(4\pi)^6 M_{\rm Max}^2} F_{3}(r_{N_{\ell''}}, r_{S_{1/3}}, r_{\delta_{1/3}}) ,\ 
{\cal O}\equiv
 \left[\begin{array}{ccc} 
1 &0 & 0 \\
0 & c_a & s_a \\
0 & -s_a & c_a \\
  \end{array}
\right]
 \left[\begin{array}{ccc} 
c_b & s_b & 0\\
 -s_b & c_b & 0\\
 0 &0 & 1 \\
  \end{array}
\right]
 \left[\begin{array}{ccc} 
c_c  & 0 & s_c\\
 0 & 1 & 0\\
 -s_c  & 0 & c_c\\
  \end{array}
\right].
\end{align}
Here $V_{MNS}$ is the Maki-Nakagawa-Sakata mixing matrix~\cite{Maki:1962mu},
${\cal O}$ is an arbitrary complex orthogonal matrix ${\cal O}^T {\cal O}=1$,
and we will adopt  the best-fit values of the neutrino oscillation data from the
global analysis in Ref.~\cite{Forero:2014bxa} and assume one
massless neutrino with normal ordering for simplicity in the numerical analysis
below.}

\subsection{Wilson coefficients for $b\to s\bar \ell \ell$ decay}  

Both anomalies in the lepton-universality violation measured
in $R_K \equiv B(B\to K\mu\mu)/B(B\to K ee)$ and the angular distributions
in $B \to K^*\mu\mu$~\cite{lhcb-2013} can be accounted for by 
the shifts in the Wilson coefficients {$C_9=-C_{10}$}.
Here we discuss the effective Hamiltonian characterizing the decay process:
\begin{align}
{\cal H}_{\rm eff}^f&=-\frac{f_{q_3 \ell_i} { f^\dag_{\ell_j q_2}} }{4m_{\delta_{4/3}}^2}
\left[(\bar s\gamma^\mu P_L b)(\bar \ell_j \gamma_\mu \ell_i) {-} (\bar s\gamma^\mu P_L b)(\bar \ell_j \gamma_\mu\gamma_5 \ell_i) \right].
\end{align}
Then one can write down the relevant Wilson coefficients as follows:
\begin{align}
(C_9)^{\ell_i \ell_j}&={-} (C_{10} )^{\ell_i \ell_j}=-\frac{1}{C_{\rm SM}}
\frac{f_{q_3 \ell_i} { f^\dag_{\ell_j q_2}} }{4m_{\delta_{4/3}}^2},\quad
C_{\rm SM}\equiv \frac{V_{tb} V^*_{ts} G_F\alpha_{\rm em}}{\sqrt2 \pi},
\end{align}
where  $\alpha_{\rm em}\approx1/137$ is the fine-structure constant, 
${\rm G_F}\approx1.17\times 10^{-5}$ GeV$^{-2}$ is the Fermi constant, 
and we focus on $i=j=2$ $(\ell_2 = \mu)$ in our case.
We can then compare them to 
the experimentally fitted values of $C_{9,10}$ for the $\mu \mu$ component 
obtained in Ref.~\cite{Descotes-Genon:2015uva} as follows:
\begin{align}
C_{9}={-} C_{10}: \ -{0.68}\; ({\rm best\ fit \ value}), 
\qquad  [{-0.85,-0.50}]\;({\rm at} \ 1\sigma), \qquad
  [{-1.22,-0.18}] \; ({\rm at} \ 3\sigma).
\label{eq:Cp910}
\end{align}

It is worthwhile to mention the LHCb measurement of $R_K = BR(B^+ \to K^+ \mu^+\mu^-)/BR(B^+ \to K^+ e^+e^-)=0.745^{+0.090}_{-0.074} \pm 0.036$, which shows a $2.6\sigma$ deviation from the SM prediction. 
The $R_K$ can simply be rewritten in terms of $X^\ell =C^{\ell}_9 -C^{\ell}_{10}$ ($\ell=e,\mu$), and its allowed region is found to be~\cite{Hiller:2003js, Hiller:2014yaa};
$ 0.7 \leq Re[X^e - X^\mu] \leq 1.5$,
where 
the $R_K$ data with $1\sigma$ errors are used. This constraint can be interpreted as 
\begin{align}
 -0.75 \lesssim C_9 \lesssim -0.35,\label{eq:rk-constraint}
\end{align}
where $X^e\approx0$.

{ We also note that flavor violating process $B \to K^* \ell \ell'$ and $B \to \ell \ell'$ can be induced by leptoquark exchange. The branching ratios of these processes are less than experimental upper limits~\cite{Amhis:2014hma, Dedes:2008iw} if the corresponding Wilson coefficient $C_9^{\ell \ell'}$ and $C_{10}^{\ell \ell'}$ satisfy $|C_{9}^{\ell \ell'}(C_{10}^{\ell \ell'})| \lesssim 1.0$~\cite{Crivellin:2015era}. Thus experimental constraints can be satisfied while achieving the best fit value of $C_9^{\mu \mu}$. Therefore we omit further discussion of these processes.}


\section{Various Constraints}

\subsection{LFVs and FCNCs at tree level}
Leptoquark models usually induce LFVs and FCNCs at tree level.
In our case, several processes can be induced from the term containing $f_{q_i\ell_j}$
in the Lagrangian.  Their contributions to the processes can be estimated 
in terms of the relevant coefficients 
of the effective Hamiltonian as~\cite{Carpentier:2010ue}
\begin{align}
({\cal H}_{\rm eff})_{ijkn}^{\bar\ell\ell\bar d d}
&=-\frac{{f_{q_k \ell_j} f_{\ell_i q_n}^\dag} }{{2} m_{\delta_{4/3}}^2} 
(\bar\ell_i\gamma^\mu P_L \ell_j)(\bar d_k\gamma_\mu P_L d_n)
\equiv
C_{LL}^{\bar\ell\ell\bar d d} (\bar\ell_i\gamma^\mu P_L \ell_j)(\bar d_k\gamma_\mu P_L d_n), \label{eq:hami1}\\
({\cal H}_{\rm eff})_{ijkn}^{\bar\ell\ell\bar uu}
&=
-\frac{{f_{q_k \ell_j} f_{\ell_i q_n}^\dag} }{2 m_{\delta_{1/3}}}
(\bar\ell_i\gamma^\mu P_L \ell_j)(\bar u_k\gamma_\mu P_L u_n)
\equiv C_{LL}^{\bar\ell\ell\bar uu}(\bar\ell_i\gamma^\mu P_L \ell_j)(\bar u_k\gamma_\mu P_L u_n), \label{eq:hami2}
\\
({\cal H}_{\rm eff})_{ijkn}^{\bar\nu\nu\bar qq}
&=
-\frac{{f_{q_k \ell_j} f_{\ell_i q_n}^\dag} }{2 m_{\delta_{1/3}}}
(\bar\nu_i\gamma^\mu P_L \nu_j)(\bar q_k\gamma_\mu P_L q_n)
\equiv C_{LL}^{\bar\nu\nu\bar qq}(\bar\nu_i\gamma^\mu P_L \nu_j)(\bar q_k\gamma_\mu P_L q_n),\label{eq:hami3},
\end{align}
where  
each of the experimental bounds is summarized in Ref.~\cite{KNO-1lp}. 

{\it $B_{d/s}\to\mu^+\mu^-$ measurements}: Recent experiments
CMS~\cite{Chatrchyan:2013bka} and LHCb~\cite{Aaij:2013aka} reported
the branching fractions of $B(B_s\to\mu^+\mu^-)$ and
$B(B_d\to\mu^+\mu^-)$, which can place useful bounds on new physics.
The bounds on the coefficients of the effective Hamiltonian in
Eq.~(\ref{eq:hami1})~\cite{Sahoo:2015wya} are given by
\begin{align}
\label{eq:Bsmumu}
& B(B_s\to\mu^+\mu^-):\quad 0\lesssim |C_{LL}^{\bar\mu\mu\bar sb}|\lesssim5\times 10^{-9}\ {\rm GeV}^{-2},\\
\label{eq:Bdmumu}
& B(B_d\to\mu^+\mu^-):\quad 1.5\times 10^{-9} \ {\rm GeV}^{-2}
\lesssim |C_{LL}^{\bar\mu\mu\bar db}|\lesssim3.9\times 10^{-9}\ {\rm GeV}^{-2}.
\end{align}
 The other modes are also given by
\begin{align}
& B(B_s\to e^+ e^-):\quad 
 |C_{LL}^{\bar ee \bar sb}|\lesssim2.54\times 10^{-5}\ {\rm GeV}^{-2},\\
& B(B_d\to e^+ e^-):\quad 
 |C_{LL}^{\bar ee\bar db}|\lesssim1.73\times 10^{-5}\ {\rm GeV}^{-2},\\
 & B(B_s\to \tau^+\tau^-):\quad 
 |C_{LL}^{\bar \tau\tau \bar sb}|\lesssim1.2\times 10^{-8}\ {\rm GeV}^{-2},\\
& B(B_d\to \tau^+ \tau^-):\quad 
 |C_{LL}^{\bar \tau\tau\bar db}|\lesssim1.28\times 10^{-6}\ {\rm GeV}^{-2}.
\end{align}

\subsection{ LFVs and FCNCs at the one-loop level}
\label{lfv-lu}
{\it LFVs at one-loop level}:
Some processes induced at the one-loop level could give stringent constraints.
In our case, $\ell_a\to\ell_b\gamma$ processes arise from the second term in 
Eq.~(\ref{eq:Yukawa}) via one-loop diagrams, and
the  branching ratio is given by
\begin{align}
 B(\ell_a\to\ell_b \gamma)
=
\frac{48\pi^3 C_a\alpha_{\rm em}}{{\rm G_F^2} m_{\ell_a}^2 }(|(a_R)_{ab}|^2+|(a_L)_{ab}|^2),
\end{align}
where $m_{a(b)}$ is the mass for the charged-lepton eigenstate, 
$C_{a}\approx(1,1/5)$ for ($a=\mu,\tau$), and  $a_{L(R)}$  is simply given by
\begin{align}
&(a_R)_{ab} \approx
{ \sum_{i=1}^3\frac{ f^\dag_{\ell_b q_i} f_{q_i \ell_a} m_{\ell_a} }{4(4\pi)^2}}
\left[\frac{1}{m_{\delta_{4/3}}^2} - \frac{1}{2m_{\delta_{1/3}}^2} \right],\quad
(a_L)_{ab} \approx
{  \sum_{i=1}^3\frac{f^\dag_{\ell_b q_i} f_{q_i\ell_a} m_{\ell_b} }{4(4\pi)^2}}
\left[\frac{1}{m_{\delta_{4/3}}^2} - \frac{1}{2m_{\delta_{1/3}}^2} \right],
\label{eq:g-2}
\end{align} 
{where we have taken the massless limit of the SM quarks inside the loop, because these masses should be tiny comparing to the leptoquark masses as we will discuss later. }
Then the current experimental upper bounds are given 
by~\cite{TheMEG:2016wtm, Adam:2013mnn}
  \begin{align}
  B(\mu\rightarrow e\gamma) &\leq4.2\times10^{-13},\quad
B(\tau\rightarrow \mu\gamma)\leq4.4\times10^{-8},
\quad   B(\tau\rightarrow e\gamma) \leq3.3\times10^{-8}~.
 \label{expLFV}
 \end{align}
 
 {\it Muon $g-2$}:
The muon anomalous magnetic moment is obtained by 
$\Delta a_\mu\approx -m_\mu[a_L+a_R]_{\mu\mu}$ in  Eq.~(\ref{eq:g-2}).
Experimentally, it has been measured with 
a high precision, and its deviation from the SM prediction is 
$\Delta a_\mu={\cal O}(10^{-9})$~\cite{g-2}.

 {\it FCNCs}: The term containing $g_{\ell_iq_j}$ in Eq.(~\ref{eq:Yukawa})
gives nonzero contributions to $b\to s\gamma$, and $K^0-\bar K^0$, and 
$B_d^0-\bar B_d^0$ mixings through the one-loop box diagrams.

 {\it $ B(b\to s\gamma)$}:
The (partial) decay rate of $b\to s\gamma$ through the box diagram is given by
\begin{align}
&\Gamma(b\to s\gamma)\approx \frac{\alpha_{em} m_b^5}
{442368\pi^4}\nn\\
&
\left|
\frac{
{g^\dag_{q_3 \ell_a} g_{\ell_a q_2}} \left[
m_{S_{1/3}}^6 - 6m_{S_{1/3}}^4 M_{N_{\ell_a}}^2 +3m_{S_{1/3}}^2 M_{N_a}^4+2M_{N_{\ell_a}}^6+12 6m_{S_{1/3}}^2 M_{N_{\ell_a}}^4\ln\left(\frac{m_{S_{1/3}}}{M_{N_{\ell_a}}}\right)  \right] }
{(m_{S_{1/3}}^2-M_{N_{\ell_a}}^2)^4}
\right|^2,
\end{align}
 then the branching ratio is given by 
 \begin{align}
 B(b\to s\gamma)\approx
&\frac{\Gamma(b\to s\gamma)}{\Gamma_{tot.}} \lesssim 3.29\times 10^{-4}\;,
\end{align}
 where  $\Gamma_{tot.}\approx 4.02\times10^{-13}$ GeV is the 
total decay width of the bottom quark, and the value on the right-handed side 
is the experimental upper bound~\cite{Lees:2012wg}.

 {\it $Q-\bar Q$ mixing}:   
The forms of $K^0-\bar K^0$, $B_d^0-\bar B_d^0$, and $D^0-\bar D^0$ mixings are, respectively, given by
\begin{align}
\Delta m_K&\approx
\frac{4} {(4\pi)^2}
\sum_{i,j=1}^3
\left[g_{\ell_i q_1} g^\dag_{q_2 \ell_i}g^\dag_{q_2\ell_j} g_{\ell_j q_1} F^K_{box}[N_{\ell_i},N_{\ell_j},S_{1/3}]\right.\label{eq:kk}\\
&+\left.
f_{\ell_i q_1}^\dag f_{q_2 \ell_i}f_{q_2\ell_j} f^\dag_{\ell_j q_1}
\left( \frac{F^K_{box}[{\nu}_i,{\nu}_j,\delta_{1/3}]}4+F^K_{box}[{\ell}_i,{\ell}_j,\delta_{4/3}]\right)
 \right] 
\lesssim 3.48\times10^{-15}[{\rm GeV}],\nn\\
\Delta m_{B_d}&\approx
\frac{4}{(4\pi)^2}
\sum_{i,j=1}^3
\left[g_{\ell_i q_3} g^\dag_{q_1\ell_i}g^\dag_{q_1\ell_j} g_{\ell_j q_3} F^B_{box}[N_{\ell_i},N_{\ell_j},S_{1/3}]\right.\label{eq:bb}\\
&+\left.
f_{\ell_i q_1}^\dag f_{q_3\ell_i}f_{q_3\ell_j} f^\dag_{\ell_j q_1}
\left( \frac{F^{{B}}_{box}[{\nu}_i,{\nu}_j,\delta_{1/3}]}4+F^{B}_{box}[{\ell}_i,{\ell}_j,\delta_{{4}/3}]\right)
 \right]
  \lesssim 3.36\times10^{-13} [{\rm GeV}],\nn\\
\Delta m_D&\approx
\frac{{4}}{(4\pi)^2}
\sum_{i,j=1}^3
f_{\ell_i q_1}^\dag f_{q_2 \ell_i}f_{q_2\ell_j} f^\dag_{\ell_j q_1} 
{
\left( \frac{F^D_{box}[\ell_i,\ell_j,\delta_{1/3}]}4+F^D_{box}[\nu_i,\nu_j,\delta_{2/3}]\right)
}
 \lesssim 6.25\times10^{-15}[{\rm GeV}],\label{eq:bb}\\
&F^Q_{box}(x,y,z)
=
\frac{5m_Q f_Q^2}{24}\left(\frac{m_Q}{m_{q}+m_{q'}}\right)^2
\int \frac{\delta(1-a-b-c-d)dadbdcdd}{ {a m_x^2+b m_y^2+(c+d) m_z^2}},
\end{align}

where $(q,q')$ are respectively $(d,s)$ for $K$,  $(b,d)$ for $B_d$, and  $(u,c)$ for $D$, each of the last inequalities of Eqs.(\ref{eq:kk}, \ref{eq:bb})
represents the upper bound on the experimental values \cite{pdg}, and
$f_K\approx0.156$ GeV, $f_B\approx0.191$ GeV, $f_{D}\approx0.212$ GeV, $m_K\approx0.498$ GeV,
and $m_B\approx5.280$ GeV, and  $m_{D}\approx 1.865$ GeV.
\footnote{Since we assume that one of the neutrino masses to be zero with
normal ordering that leads to the 
{first column in $g$ to be almost zero, i.e.,
{$g_{\ell_1q_1,\ell_1q_2,\ell_1q_3}\approx 0$},} and so these constraints can easily be evaded. }

\subsection{ Oblique parameters} 
Since $\Delta$ is a triplet under $SU(2)_L$ gauge symmetry, we need to 
take into account the constraints from the oblique parameters $S$, $T$ and $U$. 
Here we focus on the new physics contributions to the $T$ and $S$ parameter, 
$\Delta T$ and $\Delta S$, and the formulas are given by
\begin{align}
\Delta S=\frac{1}{9\pi}\ln\left[\frac{m_{\delta_{2/3}}^2}{m_{\delta_{4/3}}^2}\right],
\quad
\Delta T=\frac{16\pi}{s_{tw}^2m_Z^2}[\Pi_{\pm}(0)-\Pi_{33}(0)],\label{eq:st}
\end{align}
where $s_{tw}\approx0.22$ is the Weinberg angle, $m_Z$ is the $Z$-boson mass,
and $\Pi_{\pm}(0)$, $\Pi_{33}(0)$ are given in Appendix A.
The experimental bounds are given by \cite{pdg}
\begin{align}
(0.05 - 0.09) \le \Delta S \le (0.05 + 0.09), \quad (0.08 - 0.07) \le \Delta T \le (0.08 + 0.07).
 \end{align}
In order to avoid these constraints,
we simply take  $m_{\delta_{4/3}}= m_{\delta_{2/3}}= m_{\delta_{1/3}}$ in the
numerical analysis, because in such a degenerate case $\Delta S=\Delta T=0$.

\subsection{Dark Matter}
Here we identify $N_{\ell_1}$ as the DM candidate \cite{seto}, and denote 
its mass to be  $M_{N_{\ell_1}}\equiv M_X$.
The DM annihilation cross section is $p$-wave dominant and the dark matter
particles annihilate into the down-type quarks, via the process
{$N_{\ell_1} N_{\ell_1} \to d_j \bar d_k$} with a $S_{1/3}$ exchange with the couplings
{$g_{\ell_1 q_j}$ and $g^\dagger_{q_k\ell_1}$}.
The relic density is simply given by
\begin{align}
\Omega h^2\approx \frac{4.28\times10^{9} x_f^2 }
{12 \sqrt{g^*} M_P b_{\rm eff}},\quad b_{\rm eff}
\simeq \frac{|(g g^\dag)_{{\ell_1\ell_1}}|^2}{64 \pi}  
\frac{M_X^2 (m_{S_{1/3}}^4 + M_X^4)}{(m_{S_{1/3}}^2+M_X^2)^4} \;,
\end{align}
where $g^*\approx100$, $M_P\approx 1.22\times 10^{19}$, $x_f\approx25$.
Note that the $s$-wave contribution is suppressed since it is 
proportional to the square of the down-type quark mass.
In our numerical analysis below, we use the current experimental 
range for the relic density: $0.11\le \Omega h^2\le 0.13$~\cite{Ade:2013zuv}.

\section{Numerical analysis \label{sec:numerical}}

All relevant formulas have been derived in the last two sections,
we are going to perform numerical analysis and search for allowed parameter 
space, which satisfies all the constraints that we have discussed above.
We prepare {\it 10 million} random sampling points
for the relevant input parameters as follows:
{
\begin{align}
& m_{\delta_{1/3}} \in [1.2\times M_X\,, 5000\,]\text{GeV},\
m_{S_{1/3}}, \in [1.2\times M_X\,, {5000}\,]\text{GeV},
\nn\\&
M_{N_{\ell_1}}{(\equiv M_X)} \in [0\,, 3000\,]\text{GeV},\
M_{N_{\ell_2}}  \in [1.2 M_X\,, {7500}\,]\text{GeV},
\
M_{N_{\ell_3}}  \in [M_{N_{\ell_2}}\,, {10000}\,]\text{GeV},\nn\\ &
[a,b,c]\in (2\pi)\times[-(1+i),1+i],\ f'\in[-{3},0\if0\ln{4\pi}\fi],
\label{range_scanning}
\end{align}
}
where we fix $\lambda_0=4\pi$, 
$m_{\delta_{4/3}}\approx m_{\delta_{2/3}}\approx m_{\delta_{1/3}}$ and 
define $f_{q_i\ell_j}\equiv (\pm1)\times10^{f'_{q_i\ell_j}}$.
Notice that $f'_{q_2\ell_2}$ and $f'_{q_3\ell_2}$, which are taken to be large, 
are directly related to the requirements of $C_{9(10)}$.
The lower limit of the mass $m_{S_{1/3}}$ is taken to be $1.2 M_X$ in order 
to avoid the coannihilation region ($m_{S_{(1/3)}} \approx M_X $) so as
to satisfy the relic density of the DM.
After scanning, we find {{207}} parameter sets, which can accommodate the
neutrino oscillation data and satisfy all the constraints.

\begin{figure}[tb]
\begin{center}
\includegraphics[width=65mm]{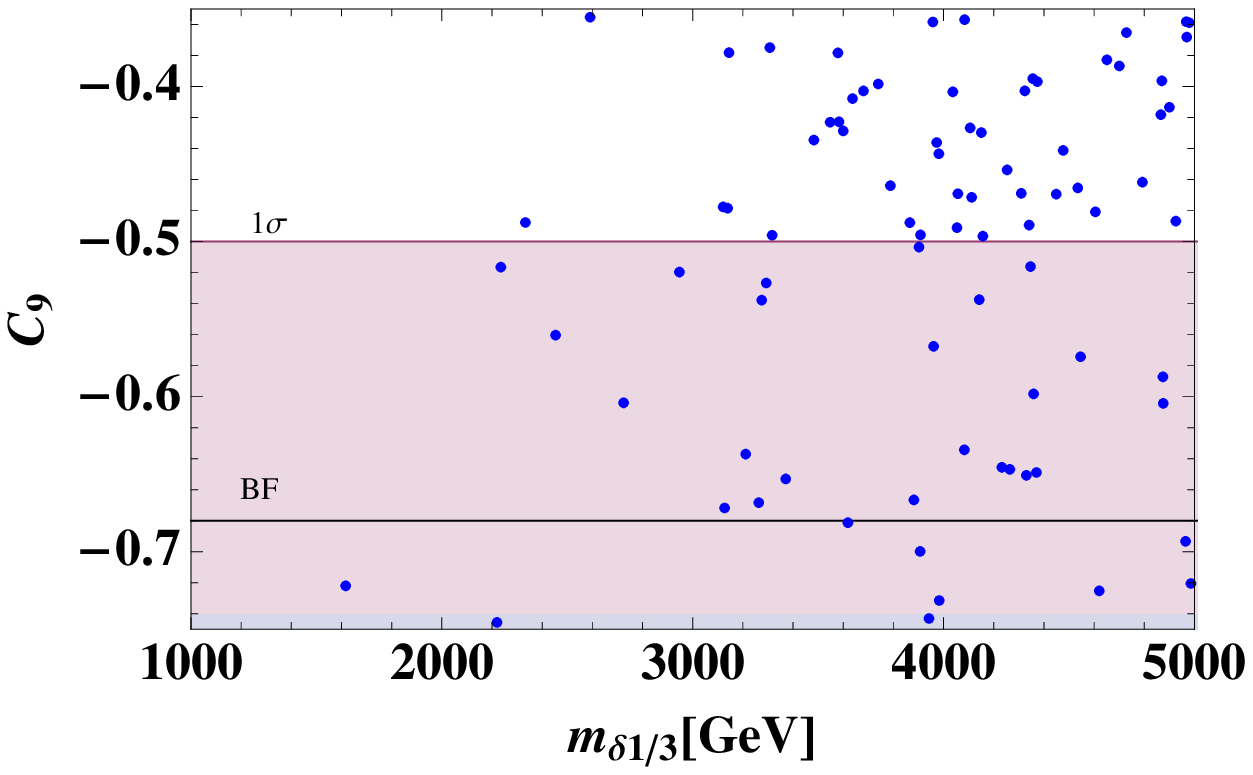}
\includegraphics[width=65mm]{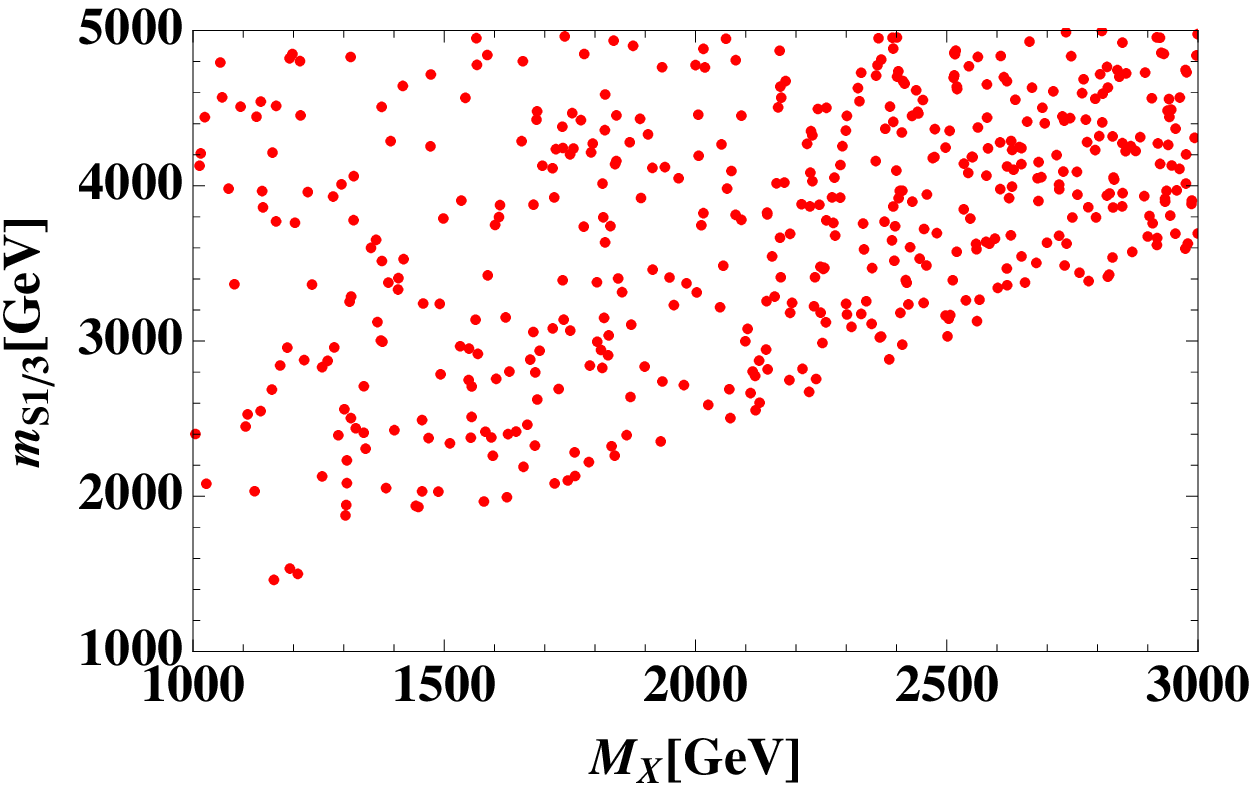}
\caption{{
Scattering plots of the allowed parameter space sets
in the plane of {$m_{\delta_{1/3}}$-$(C_{9})$} in the left
panel; and in the plane of $M_X$-$m_{S_{1/3}}$ in the right panel, {where $m_{\delta_{4/3}}\approx m_{\delta_{2/3}}\approx m_{\delta_{1/3}}$ is assumed.}
Both points satisfy the measured relic density in $[0.11-0.13]$.
The black horizontal line in the left panel represents the best fit (BF) value of $C_9(=-0.68)$, 
 the pale-red region $ [-0.85,-0.50]$ is the one at $1\sigma$ range.
Notice here that  we have taken the range $C_9\in [-0.75,-0.50]$ to satisfy $R_K$ in Eq.(\ref{eq:rk-constraint}).
}}
\label{fig:nums}
\end{center}
\end{figure}

{
In the left plot of Fig.~\ref{fig:nums}, we show the allowed region of
 $m_{\delta_{1/3}}$ versus $C_{9}$ to satisfy the measured relic density of DM, where the black horizontal line represents the best fit (BF) value of $C_9(=-0.68)$, 
 the pale-red region $ [-0.85,-0.50]$ is the one at $1\sigma$ range.
Notice here that  we have taken the range $C_9\in [-0.75,-0.50]$ to satisfy $R_K$ in Eq.(\ref{eq:rk-constraint}).
We observe that the Wilson coefficient $C_9$ can achieve 
the maximal value of $-0.8$, 
which is in agreement with the experimentally best-fitted value 
of $C_9$.


In the right plot of Fig.~\ref{fig:nums}, we show the allowed region
in the $M_X$-$m_{S_{1/3}}$ plane to satisfy the measured relic density of DM.
It suggests that
the allowed mass range for the DM is distributed to the whole range in our parameter region.

}

{We find that lower mass bound on $\delta_{1/3}$ should be about 1.6 TeV as shown in Fig.~\ref{fg}, although
$|f_{q_1\ell_j}|(j=1-3)$ runs over all the range that we take. We discuss the possible consequence of sizable $|f_{q_1\ell_2}|$ value in collider phenomenology below.
Note also that relatively large values of $f_{q_3\ell_2}$ and $f_{q_2\ell_2}$ are required in order to obtain $|C_9| \gtrsim O(0.1)$
 while these bi-product coupling $f_{q_3\ell_2}f^\dag_{\ell_2q_2}$ should be $O(10^{-2})$ for $m_{\delta_{4/3}} \sim 2$ TeV, due to the constraint of the cLFV processes such as $\mu \to e \gamma$.  }

Another important remark is that the muon $g-2$ is induced at the 
typical value of $10^{-13}$ with a negative sign, which does not help
to explain the experimental value of $O(+10^{-9})$. Notice here
that the negative sign and the small value are due to the choice 
$m_{\delta_{4/3}}\approx m_{\delta_{2/3}}\approx m_{\delta_{1/3}}$,
which was employed to evade the constraints of $\Delta S$ and $\Delta T$.

{ One of the minimal ways to induce the positive and
  sizable muon $g-2$ is to introduce another leptoquark $\Phi_{7/6}$
  as an $SU(2)_L$ doublet with $U(1)_Y$ $7/6$
  charge~\cite{Chen:2016dip}. It does not violate the neutrino
  structure, and provides a source of muon $g-2$ such as $\bar
  Q_L\Phi_{7/6}e_R$.  Also, it gives another source of {
the Wilson coefficients $C_9=C_{10}$. }
  However it does not contribute to
  any other phenomenologies such as neutrino mass matrix. Thus we just
  mention this possibility.

}

\subsection{Collider Phenomenology}

Collider phenomenology mainly concerns the interactions of the 
$\delta_{-2/3,1/3,4/3}$ bosons and $S_{-1/3}$ boson. The interactions
involving the $\Delta$ and $S$ fields with fermions can be expanded as
\begin{eqnarray}
 - {\cal L} &=& f_{q_i\ell_j} \left[
  \overline{u^c_{L_i}} \left(  \nu_{L_j} \delta_{-2/3} 
                             -\ell_{L_j} \delta_{1/3}/\sqrt{2} \right )
+ \overline{d^c_{L_i}} \left( - \nu_{L_j} \delta_{1/3}/\sqrt{2} 
                             -\ell_{L_j} \delta_{4/3} \right ) \right ]
   \nonumber \\
 &+&  g_{\ell_i q_j} \overline{N_{R_i}} d^c_{R_j}  S_{-1/3}   + h.c. 
\end{eqnarray}

\begin{figure}[t!]
\includegraphics[width=80mm]{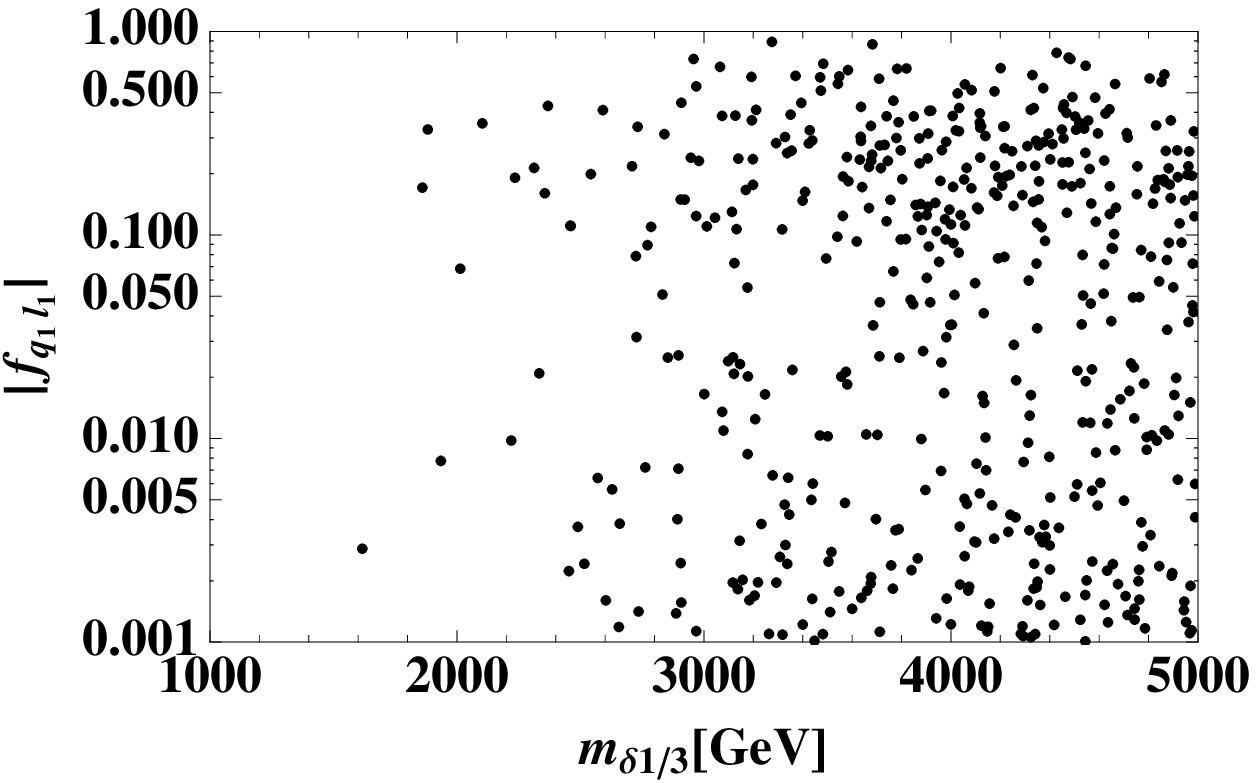}
\includegraphics[width=80mm]{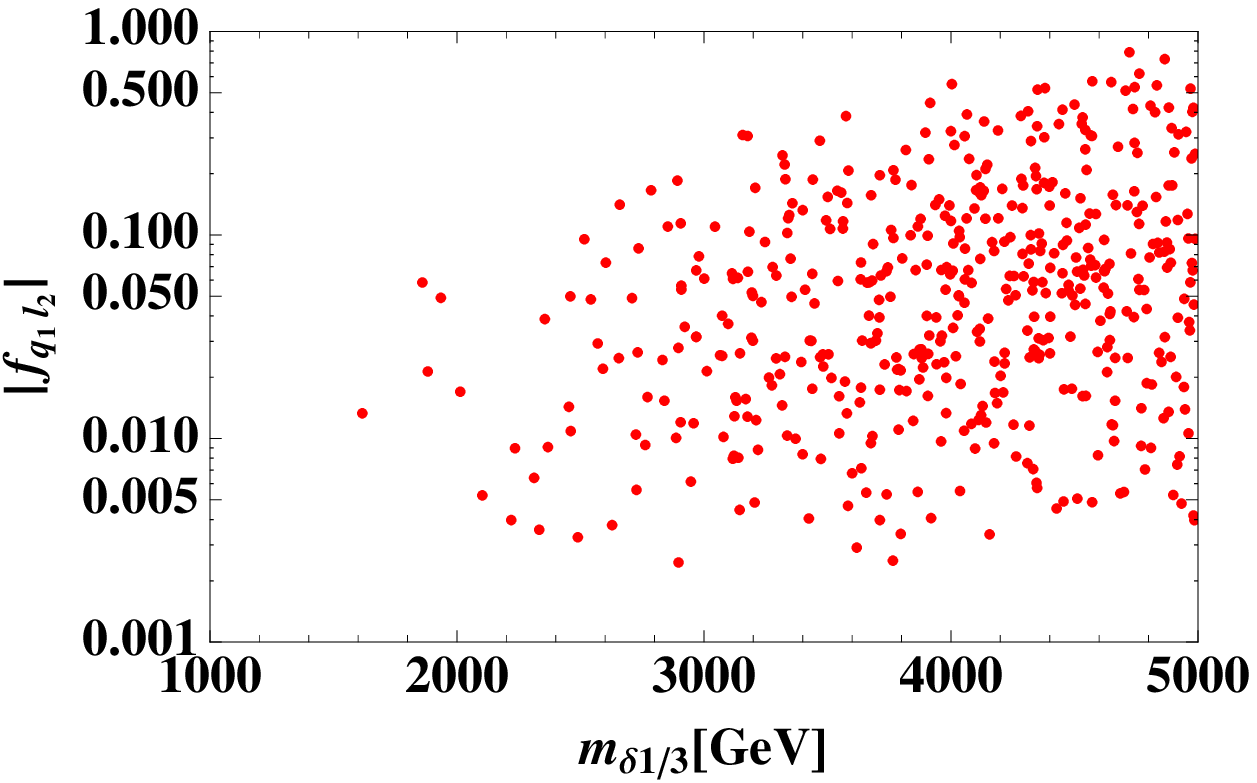}
\includegraphics[width=80mm]{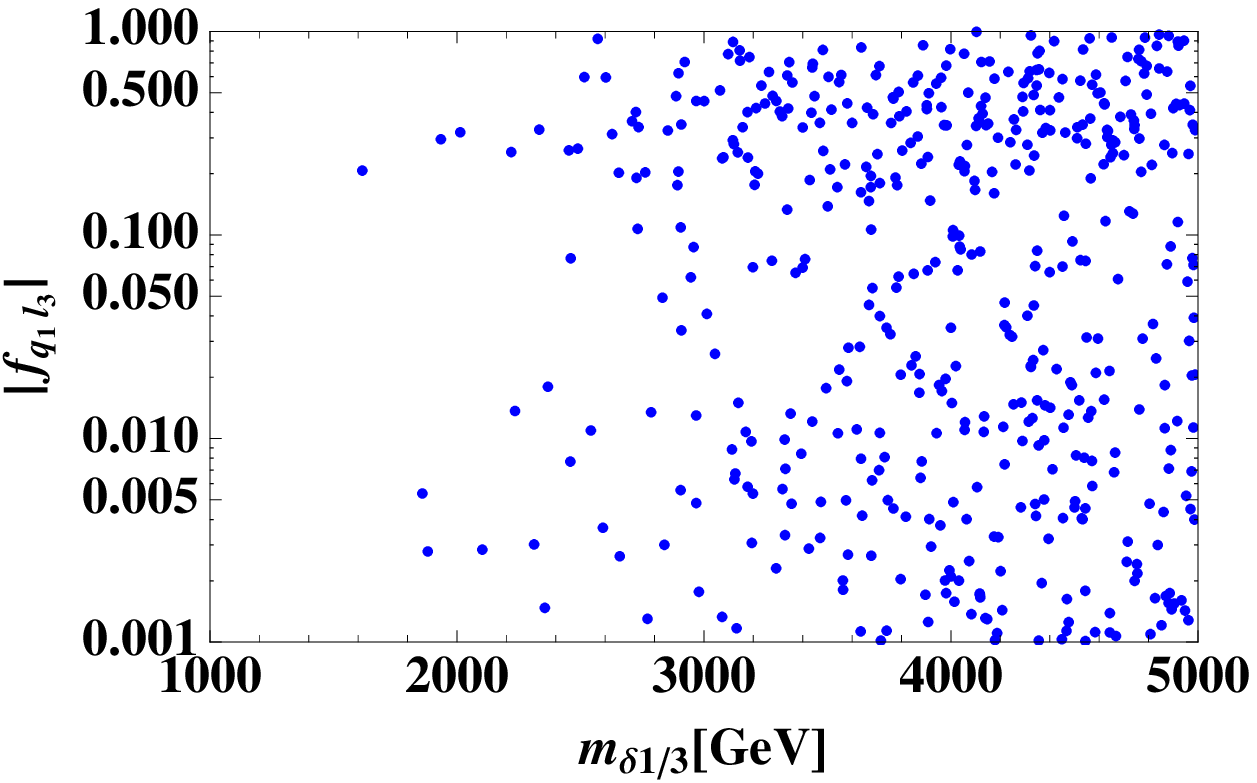}
\caption{Scatter plots of the allowed parameter space sets 
in the plane of $f_{q_1\ell_j}$ versus $m_{\delta_{4/3}},
${where $m_{\delta_{4/3}}\approx m_{\delta_{2/3}}\approx m_{\delta_{1/3}}$ is assumed.
The other components are almost the same as the one of $f_{q_1\ell_j}$.}
}
\label{fg}
\end{figure}

The $\delta$ bosons couple to a quark and a lepton (either neutrino or
charged lepton), and so they behave like leptoquarks. We first compute
the decay length of the $\delta_{-2/3}$ boson: $\delta_{-2/3} \to 
u^c_{L_i} \bar \nu_{L_j}$ summing over $i,j=1,2,3$. The result is
\begin{equation}
 \Gamma( \delta_{-2/3}) = \sum_{i,j}\; \frac{1}{16\pi} |f_{q_i\ell_j}|^2 \, m_{\delta}\;.
\end{equation}
If we take $f_{q_1\ell_2}\sim 0.1 $ to be the largest among all $f_{q_i\ell_j}$ and 
$m_{\delta}= 2$ TeV, the total width of $\delta_{-2/3} \simeq 0.4$ GeV. 
The decay is prompt.  The total widths for $\delta_{1/3}$ and $\delta_{4/3}$
are the same if their masses are the same.
The current limits for prompt leptoquarks from the LHC are roughly 1 TeV, 
depending on the search channels~\cite{lhc-lq}.  Therefore, the current
limits on $\delta$ bosons are also of order 1 TeV. The obvious production
mode for $\delta$ bosons is then the QCD pair production \cite{pair}.
The production cross section for leptoquark mass from 1 TeV to 2 TeV
goes down from 10 fb to $10^{-2}$ fb at the LHC \cite{pair}, thus 
rendering the pair production rather useless for the $\delta$ bosons
considered here, because as shown in Fig.~\ref{fg} almost all valid
parameter space points have $m_{\delta} \agt 2$ TeV.
On the other hand, one may consider the single leptoquark production in
associated with a lepton, via the subprocesses, e.g., 
$ g u \to \delta_{2/3} \bar \nu$, $g d \to \delta_{-4/3} \bar \ell$,
which involve a strong coupling and a Yukawa coupling $f_{q_i\ell_j}$ in the 
Feynman diagram \cite{singly}. This single production may be possible to 
have a larger cross section than the pair production when the Yukawa 
coupling $f_{q_i\ell_j}$ is large enough, as there is only one heavy particle
in the final state. Nevertheless, when $f_{q_i\ell_j}=e \approx 0.3$, the
cross section for $m_{\delta} = 1$ TeV is about $O(5)$ fb while
it drops down to $10^{-1}$ fb for $m_{\delta} = 2$ TeV. Since the 
cross section for single production is proportional to $|f_{q_i\ell_j}|^2$, it 
becomes $10^{-2}$ fb for $f_{q_i\ell_j}= 10^{-1}$. 
{
Unfortunately, the valid range of $|f_{q_1\ell_2}| \sim 0.02 - 0.2$ 
(See Fig.~\ref{fg}), such that 
the chance of seeing a 2 TeV $\delta$ is rather bleak.}

Another way that the $\delta$ bosons can affect is the Drell-Yan production via
a $t$-channel exchange of a $\delta$ boson, e.g., 
$u^c_{L_1} \overline{u^c_{L_1}} \to  \ell_{L_j} \bar \ell_{L_{j'}}$ via 
$\delta_{1/3}$ or
$d^c_{L_1} \overline{d^c_{L_1}} \to  \ell_{L_j} \bar \ell_{L_{j'}}$ via 
$\delta_{4/3}$. Note that $|f_{12}|$ can be as large as 
{$0.02-0.1$.}
After a Fierz transformation, the amplitude for these $t$-channel
processes can be turned into the conventional 4-fermion contact 
interactions.  We can then equate the coefficient of the amplitude to the
contact-interaction scale as 
 \[
  \frac{|f_{q_1\ell_2}|^2}{2 m_\delta^2} = \frac{4\pi}{\Lambda^2_{LL}} \;.
\]
Using the limit $\lambda_{LL}\approx 25$ TeV quoted in PDG \cite{pdg,km},
we obtain
\[
  m_\delta \agt f_{q_1\ell_2} \times 5.0 \;{\rm TeV} \;.
\]
{For $|f_{q_1\ell_2}|=0.2$ the limit on $m_\delta$  is merely 1.0 TeV.}
It is still less than the values shown in Fig.~\ref{fg}.

Now we turn to the $S^a_{\pm 1/3}$ boson, which also carries a color charge
similar to a quark. It has an $Z_2$-odd parity such that it has to be
produced in pairs. Since it is a $SU(2)_L$ singlet with a hypercharge
$-1/3$, it behaves very similar to the down-type squark $\tilde{d}$,
$\tilde{s}$, or $\tilde{b}$. The $S^a_{\pm1/3}$
so produced will decay into a down-type quark and a lighter right-handed
neutrino $N_{R_{\ell_i}}$, which is predominantly the dark matter particle
$N_{R_{\ell_1}}$.
Therefore, QCD pair production of $S_{-1/3} S_{1/3}$ gives
a final state of 2 jets plus missing energies. Depending on the mass
difference between $S_{-1/3}$ and $N_{R_{\ell_1}}$ the jets can be very soft or 
energetic enough for detection: see Fig.~\ref{fig:nums}.
To some extent we can use the limits on the squarks if
the mass difference between $S_{-1/3}$ and $N_{R_{\ell_1}}$ is large enough.
For the first two generations the limits on $\tilde{q}$ is about 
1.3 TeV \cite{atlas-q}, while for the third
generation the limit on sbottom is about 800 GeV \cite{atlas-b}.
{
Therefore, if we look back at the right panel of 
Fig.~\ref{fig:nums}, such collider limits have no effect on the valid points.
The LHC coverage at Run II can be up to about $2-3$ TeV. }

\section{Conclusions}
We have investigated a variation of the original KNT model 
with the scalar sector replaced by 
a colored $SU(2)_L$-triplet field and a colored  $SU(2)_L$-singlet field.
The model itself can afford some parameter space in accommodating
all the neutrino oscillation data and satisfying all the existing LFV
and FCNC constraints, as well as the relic density.  We have also 
successfully found the parameter sets in providing solutions to the
$b\to s\mu \bar\mu$ anomalies. 

We offer a few more comments as follows.
\begin{enumerate}

{
}

\item The solution of the colored triplet bosons to the
$b\to s\mu \bar\mu$ anomalies is similar to those by leptoquarks.
Previous works on the anomalies and leptoquarks can be found in
Ref.~\cite{lepto}

\item The contributions of the triplet  to the muon $g-2$ are
negligible compared to the experimental uncertainties.

\item The current direct limits on the $\delta$ bosons (similar to leptoquarks)
and the Drell-Yan limits are still weaker than the lower limits that 
we obtain from satisfying the $b\to s\mu\bar \mu$ anomalies, LFV and FCNC
constraints, and the DM constraint. 

\item The most interesting collider signature would be the pair 
production of $S_{-1/3} S_{1/3}$, followed by their decays into two jets
plus missing energies.

\end{enumerate}

\section*{Acknowledgments}
This work was supported by the Ministry of Science and Technology
of Taiwan under Grants No. MOST-105-2112-M-007-028-MY3.

\begin{appendix}

\section{New particle contribution to vacuum polarization diagram}
Here we summarize contributions to $\Pi_{\pm}(q^2)$ and $\Pi_{33}(q^2)$ 
in Eq.~(\ref{eq:st}) from the new particles in our model.

\noindent
{\bf Contributions to $\Pi_{\pm}(q^2)$ } \\
The one loop contributions from three-point gauge interaction are denoted by $\Pi_\pm^{XY}(q^2)$ where $X$ and $Y$ indicate particles inside loop.
They are summarized as follows;
\begin{align}
 & \Pi_{\pm}^{\delta_{1/3} \delta_{4/3}} (q^2) = \frac{2}{(4\pi)^2} G(q^2, m_{\delta_{1/3}}^2, m_{\delta_{4/3}}^2), \quad
 \Pi_{\pm}^{\delta_{1/3} \delta_{2/3}} (q^2) = \frac{2}{(4\pi)^2} 
G(q^2, m_{\delta_{1/3}}^2, m_{\delta_{2/3}}^2), 
\end{align}
where 
\begin{align}
&  G(q^2, m_P^2, m_Q^2) =  \int dx dy \delta (1-x-y) \Delta_{PQ} [\Upsilon+1 - \ln \Delta_{PQ}], \nonumber \\
 & \Delta_{PQ} = -q^2 x(1-x) + x m_{P}^2 + y m_{Q}^2, \quad \Upsilon = \frac{2}{\epsilon} - \gamma - \ln (4 \pi).
\end{align}
The one loop contributions from four-point gauge interaction are denoted by $\Pi_\pm^{X}(q^2)$ where $X$ indicates a particle inside loop.
They are summarized as follows;
\begin{align}
& \Pi_\pm^{\delta_{1/3}}(q^2) = - \frac{2}{(4\pi)^2} 
H(m_{\delta_{1/3}}^2), \quad
 \Pi_\pm^{\delta_{2(4)/3}}(q^2) = - \frac{1}{(4\pi)^2} H(m_{\delta_{2(4)/3}}^2),
\end{align}
where
\begin{equation}
H(m_P^2) = m_P^2 [\Upsilon + 1 - \ln m_P^2].
\end{equation}

\noindent
{\bf Contribution to $\Pi_{33}(q^2)$ } \\
The one loop contributions from three-point gauge interaction are denoted by $\Pi_{33}^{XY}(q^2)$ where $X$ and $Y$ indicate particles inside loop.
They are summarized as follows;
\begin{align}
& \Pi_{33}^{\delta_{2/3} \delta_{2/3}} = \frac{2}{(4\pi)^2}
G(q^2, m_{\delta_{2/3}}^2, m_{\delta_{2/3}}^2), \quad
 \Pi_{33}^{\delta_{4/3} \delta_{4/3}} = \frac{2}{(4\pi)^2} 
G(q^2, m_{\delta_{4/3}}^2, m_{\delta_{4/3}}^2). 
\end{align}
The one loop contributions from four-point gauge interaction are denoted by $\Pi_{33,3Q,QQ}^{X}(q^2)$ where $X$ indicates a particle inside loop.
They are summarized as follows;
\begin{align}
& \Pi_{33}^{\delta_{1/3}} = - \frac{2}{(4\pi)^2} H( m_{\delta_{1/3}}^2), \quad
 \Pi_{33}^{\delta_{4/3}} = - \frac{2}{(4\pi)^2} H( m_{\delta_{4/3}}^2). 
\end{align}

\end{appendix}

\end{document}